\documentclass[runningheads]{llncs}
\usepackage{graphicx}
\usepackage{multirow}
\usepackage{setspace}
\usepackage{float}
\usepackage{booktabs}
\usepackage{graphicx}
\usepackage{tabularx}
\usepackage{tabu}
\usepackage{array} 
\usepackage{threeparttable} 
\usepackage{makecell}
\usepackage{cite} 
\usepackage{amsmath} 

\begin{document}
\title{MusiCoder: A Universal Music-Acoustic Encoder Based on Transformers\thanks{Supported by Alibaba Group, and Key Laboratory of Design Intelligence and Digital Creativity of Zhejiang Province, Zhejiang University}}
\titlerunning{MusiCoder}
%
\author{Yilun Zhao\inst{1,2*}
\and
Jia Guo\inst{2}
}
\authorrunning{Y. Zhao, J. Guo}
%
\institute{Zhejiang University/University of Illinois at Urbana-Champaign Institute, Zhejiang University, Haining, China \\
\email{zhaoyilun@zju.edu.cn}\\
\and
YouKu Cognitive and Intelligent Lab, Alibaba Group, Hangzhou, China\\
\email{\{yilun.zyl, gj243069\}@alibaba-inc.com}}
\maketitle              
\begin{abstract}
Music annotation has always been one of the critical topics in the field of Music Information Retrieval (MIR). Traditional models use supervised learning for music annotation tasks. However, as supervised machine learning approaches increase in complexity, the increasing need for more annotated training data can often not be matched with available data. 
In this paper, a new self-supervised music acoustic representation learning approach named MusiCoder is proposed. Inspired by the success of BERT, MusiCoder builds upon the architecture of self-attention bidirectional transformers. 
Two pre-training objectives, including Contiguous Frames Masking (CFM) and Contiguous Channels Masking (CCM), are designed to adapt BERT-like masked reconstruction pre-training to continuous acoustic frame domain. 
The performance of MusiCoder is evaluated in two downstream music annotation tasks. The results show that MusiCoder outperforms the state-of-the-art models in both music genre classification and auto-tagging tasks. The effectiveness of MusiCoder indicates a great potential of a new self-supervised learning approach to understand music: first apply masked reconstruction tasks to pre-train a transformer-based model with massive unlabeled music acoustic data, and then finetune the model on specific downstream tasks with labeled data.
\end{abstract}
\section{Introduction}
The amount of music has been growing rapidly over the past decades. As an effective measure for utilizing massive music data, automatically assigning one music clip a set of relevant tags, providing high-level descriptions about the music clip such as genre, emotion, theme, are of great significance in MIR community \cite{recommendation_2, zhang2018pmemo}. Some researchers have applied several supervised learning models~\cite{ghosal2018music, faresnet, samplecnn, vqvae}, which are trained on human-annotated music data. However, the performance of supervised learning method are likely to be limited by the size of labeled dataset, which is expensive and time consuming to collect. 

Recently, self-supervised pre-training models \cite{speech_xlnet, ling2020deep, wang2020unsupervised, mockingjay}, especially BERT, dominate Natural Language Processing (NLP) community. BERT proposes a Masked Language Model (MLM) pre-training objective, which can learn a powerful language representation by reconstructing the masked input sequences in pre-training stage. The intuition behind this design is that a model available to recover the missing content should have learned a good contextual representation. In particular, BERT and its variants~\cite{roberta, xlnet, ernie} have reached significant improvements on various NLP benchmark tasks~\cite{glue}. 
Compared with the text domain whose inputs are discrete word tokens, in acoustics domain, the inputs are usually multi-dimensional feature vectors (e.g., energy in multiple frequency bands) of each frame, which are continuous and smoothly changed over time. Therefore, some particular designs need to be introduced to bridge the gaps between discrete text and contiguous acoustic frames. 
We are the first to apply the idea of masked reconstruction pre-training to the continuous music acoustic domain. In this paper, a new self-supervised pre-training scheme called MusiCoder is proposed, which can learn a powerful acoustic music representations through reconstructing masked acoustic frame sequence in pre-training stage.

Our contributions can be summarized as:
\begin{itemize}
\item[1.] We present a new self-supervised pre-training model named MusiCoder. MusiCoder builds upon the structure of multi-layer bidirectional self-attention transformers. Rather than relying on massive human-labeled data, MusiCoder can learn a powerful music representation from unlabeled music acoustic data, which is much easier to collect.
\item[2.] The reconstruction procedure of BERT-like model is adapted from classiﬁcation task to regression task. In other word, MusiCoder can reconstruct continuous acoustic frames directly, which avoids an extra transformation from continuous frames to discrete word tokens before pre-training. 
\item[3.] Two pre-training objectives, including Contiguous Frames Masking (CFM) and Contiguous Channels Masking (CCM), are proposed to pre-train MusiCoder. The ablation study shows that both CFM and CCM objectives can effectively improve the performance of MusiCoder pre-training.
\item[4.] The MusiCoder is evaluated on two downstream tasks: GTZAN music genre classification and MTG-Jamendo music auto-tagging. And MusiCoder outperforms the SOTA model in both tasks. The success of MusiCoder indicates a great potential of applying transformer-based masked reconstruction pre-training in Music Information Retrieval (MIR) field.
\end{itemize}

\begin{figure}[!h]
    \centering
    \includegraphics[width=1\textwidth]{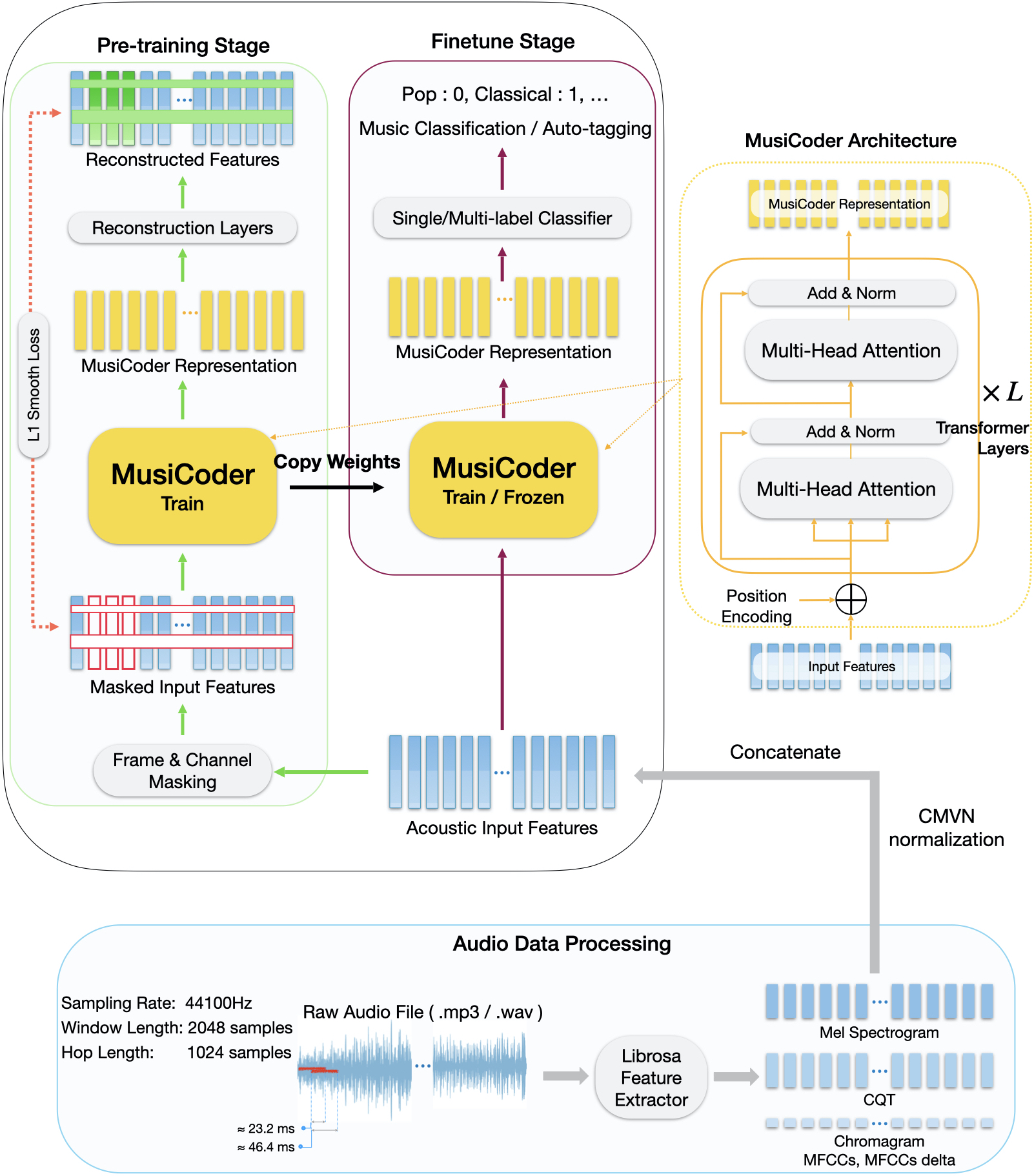}
    \caption{System overview of the MusiCoder model}
    \label{model_structure} 
    \vspace{-0.5cm}
\end{figure}
\section{Related Work}
In the past few years, pre-training models and self-supervised representation learning have achieved great success in NLP community. Huge amount of self-supervised pre-training models based on multi-layer self-attention transformers~\cite{attention}, such as BERT~\cite{bert}, GPT~\cite{gpt}, XLNet~\cite{xlnet}, Electra~\cite{electra} are proposed. Among them, BERT is perhaps the most classic and popular one due to its simplicity and outstanding performance. Specifically, BERT is designed to reconstruct the masked input sequences in pre-training stage. Through reconstructing the missing content from a given masked sequence, the model can learn a powerful contextual representation. 

More recently, the success of BERT in NLP community draws the attention of researchers in acoustic signal processing field. Some pioneering works \cite{vq_wav2vec, speech_xlnet, ling2020deep, wang2020unsupervised, mockingjay} have shown the effectiveness of adapting BERT to Automatic Speech Recognition (ASR) research. Specifically, they design some specific pre-training objectives to bridge the gaps between discrete text and contiguous acoustic frames. 
In vq-wav2vec~\cite{vq_wav2vec}, input speech audio is first discretized to a K-way quantized embedding space by learning discrete representation from audio samples. However, the quantization process requires massive computing resources and is against the continuous nature of acoustic frames. Some works \cite{speech_xlnet, ling2020deep, wang2020unsupervised, mockingjay, aalbert} design a modiﬁed version of BERT to directly utilize continuous speech. 
In \cite{ling2020deep, mockingjay, aalbert}, continuous frame-level masked reconstruction is adapted in BERT-like pre-training stage. In \cite{wang2020unsupervised}, SpecAugment~\cite{specaugment} is applied to mask input frames. And \cite{speech_xlnet} learns by reconstructing from shuffled acoustic frame orders rather than masked frames.

As for MIR community, representation learning has been popular for many years. Several convolutional neural networks (CNNs) based supervised methods \cite{choi2017transfer, ghosal2018music, faresnet, samplecnn, vqvae} are proposed in music understanding tasks. They usually employ variant depth of convolutional layers on Mel-spectrogram based representations or raw waveform signals of the music to learn effective music representation, and append fully connected layers to predict relevant annotation like music genres, tags. However, training such CNN-based models usually requires massive human-annotated data. And in \cite{attack1, attack2}, researchers show that compared with supervised learning methods, using self-supervision on unlabeled data can significantly improve the robustness of the model.
Recently, the self-attention transformer has shown promising results in symbolic music generation area. For example, Music Transformer \cite{music_transformer} and Pop Music Transformer \cite{pop_transformer} employ relative attention to capture long-term structure from music MIDI data, which can be used as discrete word tokens directly. However, compared with raw music audio, the size of existing MIDI dataset is limited. Moreover, transcription from raw audio to MIDI files is time-consuming and not accurate. In this paper, we proposed MusiCoder, a universal music-acoustic encoder based on transformers. Specifically, MusiCoder is first pre-trained on massive unlabeled music acoustic data, and then finetuned on specific downstream music annotation tasks using labeled data.
\section{MusiCoder Model}
A universal transformer-based encoder named MusiCoder is presented for music acoustic representation learning. The system overview of proposed MusiCoder is shown in Fig.~\ref{model_structure}.

\subsection{Input representation}
For each input frame $t_i$, its vector representation $x_i$ is obtained by first projecting $t_i$ linearly to hidden dimension $H_{dim}$, and then added with sinusoidal positional encoding~\cite{attention} defined as following:
\begin{equation}
\begin{aligned}
    PE_{(pos, 2i)} & = sin(pos/10000^{2i/H_{dim}}) \\
    PE_{(pos, 2i+1)} & = cos(pos/10000^{2i/H_{dim}})
\end{aligned}
\end{equation}
The positional encoding is used to inject information about the relative position of acoustic frames. The design of positional encoding makes the transformer encoder aware of the music sequence order.

\subsection{Transformer Encoder}
A multi-layer bidirectional self-attention transformer encoder~\cite{attention} is used to encode the input music acoustic frames. Specifically, a $L$-layer transformer is used to encode the input vectors $X = \left\{x_i\right\}^N_{i=1}$ as: 
\begin{equation} 
    H^l = Transformer_l(H^{l-1})
\end{equation}
where $l \in [1, L]$, $H^0 = X$ and $H^L = [h_1^L, ... , h_N^L]$. We use the hidden vector $h^L_i$ as the contextualized representation of the input token $t_i$. The architecture of transformer encoder is shown in Fig. \ref{model_structure}.

\subsection{Pre-training Objectives}
The main idea of masked reconstruction pre-training is to perturb the inputs by randomly masking tokens with some probability, and reconstruct the masked tokens at the output.
In the pre-training process, a reconstruction module, which consists of two layers of feed-forward network with GeLU activation \cite{gelu} and layer-normalization \cite{layer_normalization}, is appended to predict the masked inputs. The module uses the output of the last MusiCoder encoder layer as its input.
Moreover, two new pre-training objectives are presented to help MusiCoder learn acoustic music representation.

\subsubsection{Objective 1: Contiguous Frames Masking (CFM).}
To avoid the model exploiting local smoothness of acoustic frames, rather than only mask one span with fixed number of consecutive frames~\cite{mockingjay}, we mask several spans of consecutive frames dynamically.
Given a sequence of input frames $X= (x_{1}, x_{2}, ... ,x_{n})$, we select a subset $Y \subset X$ by iteratively sampling contiguous input frames (spans) until the masking budget (e.g., 15\% of $X$) has been spent. At each iteration, the span length is first sampled from a geometric distribution $\ell \sim Geo(p)$. Then the starting point of the masked span is randomly selected. We set $p= 0.2$, $\ell_{min}=2$ and $\ell_{max}=7$. The corresponding mean length of span is around 3.87 frames ($\approx179.6ms$). In each masked span, the frames are masked according to the following policy: 
1) replace all frames with zero in 70\% of the case. Since each dimension of input frames are normalized to have zero mean value, setting the masked value to zero is equivalent to setting it to the mean value. 
2) replace all frames with a random masking frame in 20\% of the case. 
3) keep the original frames unchanged in the rest 10\% of the case. Since MusiCoder will only receive acoustic frames without masking during inference time, policy 3) allows the model to receive real inputs during pre-training, and resolves the pretrain-fintune inconsistency problem~\cite{bert}.

\subsubsection{Objective 2: Contiguous Channels Masking (CCM).}
The intuition of channel masking is that a model available to predict the partial loss of channel information should have learned a high-level understanding along the channel axis. For log-mel spectrum and log-CQT features, a block of consecutive channels is randomly masked to zero for all time steps across the input sequence of frames. Specifically, the number of masked blocks, $n$, is first sampled from $\left\{0, 1, \dots, H\right\}$ uniformly. Then a starting channel index is sampled from $\left\{0, 1, \dots, H-n\right\}$, where $H$ is the number of total channels.

\subsubsection{Pre-training Objective Function.}
\begin{equation}
    Loss(x) = \left\{
                        \begin{array}{lr}
                        0.5\cdot x^{2} \quad\;\; if \; |x| < 1 \\
                        |x| - 0.5 \quad otherwise
                        \end{array}
                        \right.
\end{equation}  
The Huber Loss~\cite{fastrcnn_l1smooth} is used to minimize reconstruction error between masked input features and corresponding encoder output. Huber Loss is a robust L1 loss that is less sensitive to outliers. And in our preliminary experiments, we found that compared with L1 loss used in ~\cite{mockingjay}, using Huber loss will make the training process easier to converge.

\subsection{MusiCoder Model Setting}

We primarily report experimental results on two models: MusiCoderBase and MusiCoderLarge. The model settings are listed in Table \ref{model_setting}.
The number of Transformer block layers, the size of hidden vectors, the number of self-attention heads are represented as $L_{num}$, $H_{dim}$, $A_{num}$, respectively. 

\begin{table}
\centering
\caption{The proposed model settings}
\begin{tabular}{lcccc}
\toprule
               & $L_{num}$ & $H_{dim}$  & $A_{num}$     &  \#parameters  \\
\midrule
MusiCoderBase  & 4         & 768        & 12            &  29.3M\\
MusiCoderLarge & 8         & 1024       & 16            &  93.1M\\
\bottomrule
\end{tabular}
\label{model_setting}
\vspace{-1cm}
\end{table}

\section{Experiment Setup}
\subsection{Dataset Collection and Preprocess}
\begin{table}[H]
\begin{threeparttable}
\vspace{-0.6cm}
\centering
\caption{Statistics on the datasets used for pre-training and downstream tasks}
\begin{tabular}{p{2.2cm}<{\centering}p{2.6cm}<{\centering}p{1.3cm}<{\centering}p{1.8cm}<{\centering}c}
\toprule
Task                          & Datasets          & \#clips    & duration (hours)  & Description\\
\midrule
\multirow{3}{*}{Pre-training}   & Music4all         &  109.2K           & 908.7 & \multicolumn{1}{c}{--}\\
                                & FMA-large         &  106.3K           & 886.4 & \multicolumn{1}{c}{--}\\
                                & MTG-Jamendo\tnote{1} &  37.3K         & 1346.9       & \multicolumn{1}{c}{--}   \\
                              
\addlinespace[0.1cm]
Classification            & GTZAN             &  1000             & 8.3   & 100 clips for each genre\\
\addlinespace[0.1cm]
Auto-tagging                    & MTG-Jamendo\tnote{2}  & 18.4K         & 157.1             & 56 mood/theme tags\\
\bottomrule
\end{tabular}
\begin{tablenotes}
    \footnotesize
    \item[1,2] For MTG-Jamendo dataset, we removed music clips used in Auto-tagging task when pre-training. 
\end{tablenotes}
\label{dataset_info}
\end{threeparttable}
\vspace{-0.8cm}
\end{table}

As shown in Table \ref{dataset_info}, the pre-training data were aggregated from three datasets: Music4all~\cite{music4all}, FMA-Large~\cite{fma} and MTG-Jamendo~\cite{mtg_dataset}. Both Music4all and FMA-Large datasets provide 30-seconds audio clips in .mp3 format for each song. And MTG-Jamendo dataset contains 55.7K music tracks, each with a duration of more than 30s. Since the maximum time stamps of MusiCoder is set to 1600, those music tracks exceeding 35s would be cropped into several music clips, the duration of which was randomly picked from 10s to 35s.

GTZAN music genre classification \cite{gtzan} and MTG-Jamendo music auto-tagging tasks \cite{mtg_dataset} were used to evaluate the performance of finetuned MusiCoder. GTZAN consists of 1000 music clips divided into ten different genres (blues, classical, country, disco, hip-hop, jazz, metal, pop, reggae \& rock). Each genre consists of 100 music clips in .wav format with a duration of 30s. To avoid seeing any test data in downstream tasks, for pre-training data, we filtered out those music clips appearing in downstream tasks.

\textbf{Audio Preprocess.} The acoustic music analysis library, Librosa~\cite{librosa}, provides flexible ways to extract features related to timbre, harmony, and rhythm aspect of music. In our work, Librosa was used to extract the following features from a given music clip: Mel-scaled Spectrogram, Constant-Q Transform (CQT), Mel-frequency cepstral coefficients (MFCCs), MFCCs delta and Chromagram, as detailed in Table \ref{extracted_feature}. Each kind of features was extracted at the sampling rate of 44,100Hz, with a Hamming window size of 2048 samples ($\approx$ 46 ms) and a hop size of 1024 samples ($\approx$ 23 ms). The Mel Spectrogram and CQT features were transformed to log amplitude with $S^{'}=ln(10 \cdot S+\epsilon)$, where $S$, $\epsilon$ represents the feature and an extremely small number, respectively. Then Cepstral Mean and Variance Normalization (CMVN) \cite{cmvn_1, cmvn_2} were applied to the extracted features for minimizing distortion caused by noise contamination. Finally these normalized features were concatenated to a 324-dim feature, which was later used as the input of MusiCoder. 

\begin{table}[H]
\centering
\caption{Acoustic features of music extracted by Librosa}
\begin{tabular}{p{4.5cm}<{\centering}p{3.8cm}<{\centering}p{1.5cm}<{\centering}}
\toprule
Feature                 & Characteristic    & Dimension \\
\midrule
Chromagram              & Melody, Harmony          & 12  \\
MFCCs                    & Pitch                    & 20 \\
MFCCs delta              & Pitch                    & 20 \\
Mel-scaled Spectrogram  & Raw Waveform             & 128 \\
Constant-Q Transform    & Raw Waveform             & 144 \\
\bottomrule
\end{tabular}
\label{extracted_feature}
\end{table}
\vspace{-0.8cm}

\subsection{Training Setup}
All our experiments were conducted on 5 GTX 2080Ti and can be reproduced on any machine with GPU memory more than 48GBs. In pre-training stage, MusiCoderBase and MusiCoderLarge were trained with a batch size of 64 for 200k and 500k steps, respectively. We applied the Adam optimizer~\cite{adam} with $\beta_1 = 0.9$, $\beta_2=0.999$ and $\epsilon=10^{-6}$. And the learning rate were varied with warmup schedule~\cite{attention} according to the formula:
\begin{equation}
    lrate=H_{dim}^{-0.5} \cdot min(step\_num^{-0.5}, step\_num \cdot warmup\_steps^{-1.5})
\end{equation}
where $warmup\_steps$ was set as 8000. Moreover, library Apex was used to accelerate the training process and save GPU memory.

For downstream tasks, we performed an exhaustive search on the following sets of parameters. The model that performed the best on the validation set was selected. All the other training parameters remained the same as those in pre-training stage: 

\begin{table}[]
\centering
\caption{Parameter settings for downstream tasks}
\begin{tabular}{p{2.5cm}<{\centering}p{4cm}<{\centering}}
\toprule
Parameter               & Candidate Value   \\
\midrule
Batch size              & 16, 24, 32        \\
Learning Rate           & 2e-5, 3e-5, 5e-5  \\
Epoch                  & 2, 3, 4            \\
Dropout Rate            & 0.05, 0.1         \\
\bottomrule
\end{tabular}
\label{extracted_feature}
\end{table}
\vspace{-1cm}


\section{Results}
\subsection{Music Genre Classification}
\vspace{-0.6cm}
\begin{table}[H]
\centering
\caption{Results of GTZAN Music Classification task}
\begin{tabular}{p{6cm}c}
\toprule
Models        & accuracy  \\
\midrule
hand-crafted features + SVM  \cite{baniya2014audio} & 87.9\%    \\
CNN + SVM \cite{choi2017transfer}                   & 89.8\%    \\           
CNN+MLP based ensemble \cite{ghosal2018music}       & 94.2\%    \\
\textbf{MusiCoderBase}                              & \textbf{94.2\%}    \\
\textbf{MusiCoderLarge}                             & \textbf{94.3\%}    \\
Theoretical Maximum Score\cite{gtzan}               & 94.5\%    \\
\bottomrule
\end{tabular}
\label{classification}
\vspace{-0.6cm}
\end{table}
Since GTZAN dataset only contains 1000 music clips, the experiments were conducted in a ten-fold cross-validation setup. For each fold, 80, 20 songs of each genre were randomly selected and placed into the training and validation split, respectively. The ten-fold average accuracy score is shown in Table~\ref{classification}. In prevoious work, \cite{baniya2014audio} applied low-level music features and rich statistics to predict music genres. In \cite{choi2017transfer}, researchers first used a CNN based model, which was trained on music auto-tagging tasks, to extract features. These extracted features were then applied on SVM~\cite{svm} for genre classification. In \cite{ghosal2018music}, the authors trained two models: a CNN based model trained on a variety of spectral and rhythmic features, and an MLP network trained on features, which were extracted from a model for music auto-tagging task. Then these two models were combined in a majority voting ensemble setup. The authors reported the accuracy score as 94.2\%. Although some other works reported their accuracy score higher than 94.5\%, we set 94.5\% as the state-of-the-art accuracy according to the analysis in \cite{gtzan}, which demonstrates that the inherent noise (e.g., repetitions, mis-labelings, distortions of the songs) in GTZAN dataset prevents the perfect accuracy score from surpassing 94.5\%. 
In the experiment, MusiCoderBase and MusiCoderLarge achieve accuracy of 94.2\% and 94.3\%, respectively. The proposed models outperform the state-of-the-art models and achieve accuracy score close to the ideal value.

\subsection{Music Auto-Tagging}
\vspace{-0.6cm}
\begin{table}[]
\centering
\caption{Results of MTG-Jamendo Music Auto-tagging task}
\begin{tabular}{lp{6.5cm}p{1.8cm}<{\centering}p{1.8cm}<{\centering}}
\toprule
& Models                            & ROC-AUC macro     & PR-AUC macro   \\
\midrule
& VQ-VAE+CNN \cite{vqvae}         & 72.07\%           & 10.76\%   \\
& VGGish  \cite{mtg_dataset}           & 72.58\%           & 10.77\%   \\
& CRNN \cite{faresnet}         & 73.80\%           & 11.71\%   \\
& FA-ResNet \cite{faresnet}         & 75.75\%           & 14.63\%   \\
& SampleCNN (reproduced) \cite{samplecnn}  & 76.93\%           & 14.92\%  \\
& Shake-FA-ResNet \cite{faresnet}   & 77.17\%           & 14.80\%   \\
\midrule
\multirow{5}{*}{Ours} & MusiCoderBase w/o pre-training       & 77.03\%   & 15.02\% \\
& MusiCoderBase with CCM            & 81.93\%           & 19.49\%   \\
& MusiCoderBase with CFM            & 81.38\%           & 19.51\%   \\
& \textbf{MusiCoderBase with CFM+CCM} & \textbf{82.57\%} & \textbf{20.87\%} \\
& \textbf{MusiCoderLarge with CFM+CCM}  & \textbf{83.82\%} & \textbf{22.01\%}  \\
\bottomrule
\end{tabular}
\label{tagging}
\vspace{-0.6cm}
\end{table}
For the music auto-tagging task, two sets of performance measurements, ROC-AUC macro and PR-AUC macro, were applied. ROC-AUC can lead to over-optimistic scores when data are unbalanced \cite{pr_roc}. Since the music tags given in the MTG-Jamendo dataset are highly unbalanced \cite{mtg_dataset}, the PR-AUC metric was also introduced for evaluation. The MusiCoder model was compared with other state-of-the-art models competing in the challenge of MediaEval 2019: Emotion and Theme Recognition in Music Using Jamendo~\cite{mtg_dataset}. We used the same train-valid-test data splits as the challenge.
The results are shown in Table~\ref{tagging}. For VQ-VAE+CNN~\cite{vqvae}, VGGish~\cite{mtg_dataset}, CRNN~\cite{faresnet}, FA-ResNet~\cite{faresnet}, Shake-FA-ResNet~\cite{faresnet} models, we directly used the evaluation results posted in the competition leaderboard\footnote{\url{https://multimediaeval.github.io/2019-Emotion-and-Theme-Recognition-in-Music-Task/results}}. For SampleCNN~\cite{samplecnn}, we reproduced the work according to the official implementation\footnote{\url{https://github.com/tae-jun/sample-cnn}}. As the results suggest, the proposed MusiCoder model achieves new state-of-the-art results in music auto-tagging task. 

\subsubsection{Ablation Study.}
Ablation study were conducted to better understand the performance of MusiCoder. The results are also shown in Table~\ref{tagging}.
According to the experiemnt, even without pre-training, MusiCoderBase can still outperform most SOTA models, which indicates the effectiveness of transformer-based architecture.
When MusiCoderBase is pre-trained with objective CCM or CFM only, a signiﬁcant improvement over MusiCoderBase without pre-training is observed. And MusiCoderBase with CCM and CFM pre-training objectives combined achieves better results.  
The improvement indicates the effectiveness of pre-training stage. 
And it shows that the designed pre-training objectives CCM and CFM are both the key elements that drives pre-trained MusiCoder to learn a powerful music acoustic representation.
We also explore the effect of model size on downstream task accuracy. In the experiment, MusiCoderLarge outperforms MusiCoderBase, which reflects that increasing the model size of MusiCoder will lead to continual improvements.

\section{Conclusion}
In this paper, we propose MusiCoder, a universal music-acoustic encoder based on transformers. Rather than relying on massive human labeled data which is expensive and time consuming to collect, MusiCoder can learn a strong music representation from unlabeled music acoustic data.
Two new pre-training objectives Contiguous Frames Masking (CFM) and Contiguous Channel Masking (CCM) are designed to improve the pre-training stage in continuous acoustic frame domain. The effectiveness of proposed objectives is evaluated through extensive ablation studies. Moreover, MusiCoder outperforms the state-of-the-art model in music genre classification on GTZAN dataset and music auto-tagging on MTG-Jamendo dataset. 
Our work shows a great potential of adapting transformer-based masked reconstruction pre-training scheme to MIR community. 
Beyond improving the model, we plan to extend MusiCoder to other music understanding tasks (e.g., music emotion recognition, chord estimation, music segmentation). We believe the future prospects for large scale representation learning from music acoustic data look quite promising.

%
%
%
%

\bibliographystyle{splncs04}
\bibliography{references}
\end{document}